\documentclass[prl,twocolumn]{revtex4}
\usepackage{amsmath}
\usepackage{graphicx}

\begin{document}
	
	\title{Tracking the connection between disorder and energy landscape in glasses using geologically hyperaged amber}
	
	\author{E. A. A. Pogna$^{1,2}$, A. I. Chumakov$^{3,4}$, C. Ferrante$^{5,6}$, M. A. Ramos$^7$, T. Scopigno$^{5,6}$}
	
	\affiliation{
		$^1$Laboratorio NEST, CNR-INFM and Scuola Normale Superiore, Piazza San Silvestro 12, I-56127 Pisa, Italy
		$^2$Dipartimento di Fisica, Politecnico di Milano, Piazza Leonardo da Vinci 32, I-20133 Milano, Italy \\
		$^3$ESRF-The European Synchrotron Radiation Facility CS40220, F-38043, Grenoble Cedex 9, France \\
		$^4$National Research Centre "Kurchatov Institute", 123182 Moscow, Russia\\
		$^5$Dipartimento di Fisica, Universit\'a di Roma, La Sapienza, I-00185, Rome, Italy\\
		$^6$Center for Life Nano Science@Sapienza, Istituto Italiano di Tecnologia, Viale Regina Elena 291, 00161 Rome, Italy\\
		$^7$Laboratorio de Bajas Temperaturas, Departamento de F\'isica de la Materia Condensada, Condensed Matter Physics Center (IFIMAC) and Instituto Nicol\'as Cabrera, Universidad Aut\'onoma de Madrid, E-28049~Madrid,~Spain}


\begin{abstract}
Fossil amber offers the unique opportunity of investigating an amorphous material which has been exploring its energy landscape for more than 110 Myears of natural aging. By applying different x-ray scattering methods to amber before and after annealing the sample to erase its thermal history, we identify a link between the potential energy landscape and the structural and vibrational properties of glasses. We find that hyperaging induces a depletion of the vibrational density of states in the THz region, also ruling the sound dispersion and attenuation properties of the corresponding acoustic waves. Critically, this is accompanied by a densification with structural implications different in nature from that caused by hydrostatic compression. Our results, rationalized within the framework of fluctuating elasticity theory, reveal how upon approaching the bottom of the potential energy landscape (9\% decrease in the fictive temperature $T_f$) the elastic matrix becomes increasingly less disordered (6\%) and longer-range correlated (22\%).
\end{abstract}

\maketitle

Glasses are metastable systems whose microscopic properties vary depending on their thermal history and aging. The way their structure and dynamics relate to thermodynamic metastability is one of the current challenges for the glass transition understanding. From the experimental point of view, one of the major hurdles is the limited stability range normally accessible on the laboratory timescale. The glassy state is conventionally obtained by cooling the liquid below the melting point, avoiding its crystallization. The slower is the cooling rate, the lower is the energy of the basins (inherent structures) of the
potential energy landscape \cite{doi:10.1063/1.1672587,Stillinger1935} in which the system gets trapped. The stability range covered by tuning the cooling rate is
however limited: upwards by thermal conductivity and heating/cooling technology (hyperquenched glasses), downwards by crystal nucleation. Nevertheless, being metastable materials, glasses spontaneously evolve towards states of lower configurational energy and enthalpic content on a timescale dictated by the relaxation time. Accordingly, the stabilization process, or \textit{aging}, becomes increasingly slow upon lowering the temperature below the glass transition temperature $T_g$.

Amber, a fossilized resin resulting from the chemical vitrification \cite{nature_vitri} of conifers and angiosperms exudates, represents a rare example of glass which has been aging for a time as long as tens of million years (hyperaging) at temperatures well below $T_g$, experiencing extreme thermodynamic stabilization \cite{Zhao20137041,Mckenna,ramos_velocity}.
Such a long-aging history enables a unique opportunity to investigate how microscopic properties depend on energy landscape topology.

Glasses are characterized, at the microscopic scale, by the lack of long-range structural order and, at the macroscopic scale, by thermodynamical and transport properties strikingly different from those of the corresponding crystalline phase. Among them, there are the thermal conductivity plateau and a specific heat $C_p(T)$ excess over the Debye value ($\propto T^3$) located around 10 K \cite{zeller_thermal_1971-1}.
These universal anomalies can be traced back to an excess in the vibrational density of states (VDOS), over the Debye model expectations ($\propto E^2$), conventionally referred to as Boson peak (BP)\cite{binder_glassy_2005}.
Experimentally, the BP of glasses has been intensively investigated by Raman scattering \cite{amorph_sol,malinovsky1986nature,PhysRevB.67.024203,PhysRevLett.111.245502}, inelastic neutron scattering \cite{PhysRevLett.53.2316,zorn1995neutron,ramos_cp}, inelastic x-ray scattering \cite{scopigno2011vibrational}, nuclear inelastic scattering \cite{PhysRevLett.106.225501} and it has been numerically calculated by molecular dynamics simulations \cite{PhysRevLett.97.055501,schirmacher_acoustic_2007,shintani_universal_2008,marruzzo_heterogeneous_2013}.

Here, we study the evolution of the vibrational properties of the amber glass upon aging, comparing a pristine (pr) fossilized sample from El Soplao \cite{ElSoplao,ambra_qui}, Spain, aged for 110 Myears, with the corresponding ordinary glass obtained by thermal annealing and subsequent standard cooling, the annealed (an-) amber. Specifically, we combine the VDOS measurements with experimental determinations of the acoustic attenuation and phonon dispersion in the BP frequency region by means of two x-ray scattering experiments performed at the European Synchrotron Radiation Facility (ESRF): Inelastic X-ray Scattering (IXS) \cite{sette_dynamics_1998,pogna} at the beamline ID28 and Inelastic X-ray scattering with Nuclear Resonance Analysis (IXS-NRA) \cite{rufleINS} at beamline ID18, both with very high energy resolution (meV) .

\begin{figure}[!t]
\begin{center}
		\includegraphics[width=0.48\textwidth]{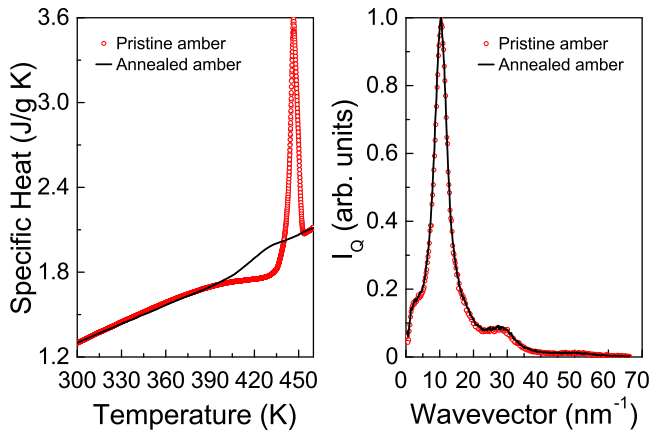}
		\caption{Left: Specific heat measured by DSC of pristine (red open dots) and annealed amber (black line). The devitrification temperature increases with the stability, $T= 447$ K for pristine amber, $T= 428$ K for the annealed. Corresponding enthalpy curves and fictive temperatures are reported in the Supporting Information; Right: X-ray diffraction pattern of pristine (red open dots) and annealed amber (black line). The intensity profile, I(Q), is normalized to the main peak.}
		\label{f:static}
\end{center}
\end{figure}

The thermodynamical stability of the pristine amber glass is characterized via differential scanning calorimetry (DSC). The specific heat $C_p$ dependence over temperature during a temperature upscan (at 10 K/min) is reported in Fig.\ref{f:static}.
The prolonged aging of pristine amber is testified by the presence of a prominent endothermic peak in the specific-heat curve around the glass-transition into supercooled liquid, which is accompanied by a strong enthalpy release ($\Delta H\approx 9$ J g$^{-1}$).

The ordinary amber glass is obtained by heating the sample above $T_g$ ($T \geq 423$ K) for 3h to erase its thermal history and then vitrifying it at the standard cooling rate of 10 K/min. Such a treatment leads to a material with significant enthalpy increase, quantified by a variation of the enthalpic fictive temperature $T_f$, i.e. the temperature at which the glass would be in equilibrium with its own liquid\cite{moy}. Following the standard procedure detailed in SI we found a fractional increase $\Delta T_f/T_f$ of about 9 $\%$ from pristine ($T_f$=379 K) to annealed ($T_f=415$ K) amber, in line with previously reported experiments on El Soplao amber \cite{perez,perez2016} and on other amber, testifying to a substantial material "rejuvenation".

Having assessed the different stability degrees of pristine and annealed amber, we investigated their structure by wide angle x-ray diffraction at the beamline ID28 of ESRF. The scattered intensity I(Q), proportional to the static structure factor S(Q), is reported in Fig.\ref{f:static}.
Despite the different thermodynamical stability, the two samples share similar gross structural features, showing a main peak centered at Q$_{p}\approx 10$ nm$^{-1}$ and two smaller, broader peaks at Q$_{p}^{'}\approx 28$ nm$^{-1}$ and Q$_{p}^{''}\approx 52$ nm$^{-1}$. Remarkably, the main peak of the pristine amber (which is a $2\%$  denser than the annealed amber \cite{perez}) can be relatively scaled on top of the main peak of an-amber by applying a $0.75\%$ shrinking factor to the Q-axis. 

This behaviour has to be contrasted with the linear dependence of the first diffraction peak on density in compressed tetrahedral glasses \cite{stone2001structure}, traced back to a reduction of the intermediate range order and average size of the network cages. The structural observation reported here, therefore, is highly suggestive of a different way (aging-induced) of achieving the densification. 

\begin{figure}[ht]
	\begin{center}
		\includegraphics[width=0.48\textwidth]{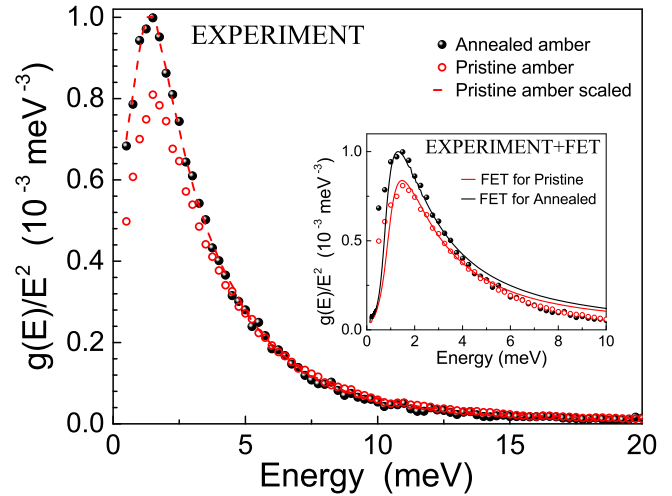}
		\caption{Boson Peak of pristine (red open dots) and annealed amber (black dots) at 30K measured by IXS-NRA. The two BP are in excellent agreement after 11\% horizontal and 26\% vertical scaling of the former (red dashed line). In the inset the two (absolute) BP have been fitted with the Fluctuating Elasticity Theory, which allows to quantify the distribution and correlation length of the elastic matrix (see text).}
		\label{f:DOS}
	\end{center}
\end{figure}
The VDOS of the two samples was measured by means of inelastic x-ray scattering with nuclear resonance analysis (IXS-NRA) \cite {rufleINS} at the nuclear resonance beamline \cite{R} ID18 of ESRF. Experimental details are discussed in SI.

The excess of vibrational density of states over the Debye expectation $g(E)/E^2$, i.e. the BP, is peaked at about $E_{BP}=1.5$ meV, in good agreement with the temperature-position of the broad maximum in the normalized specific heat $C_p/T^3$ observed in pristine and annealed amber at T$_{BP}\approx 3.4$ K \cite{perez}. According to the direct correlation $E_{BP}=(4.5-5.0)k_B T_{BP}$, based on a number of different experimental data \cite{PhysRevB.80.094303,carini}, the BP was expected at 1.3 - 1.5 meV, in excellent agreement with IXS-NRA data of Fig.\ref{f:DOS}.
In order to compare the VDOS of the two samples, we applied an horizontal and a vertical scaling to the pristine measurement to best fit the annealed curve. Remarkably, a very good overlap is found by shrinking the horizontal axis by the $11\%$ and expanding the vertical axis by the $26\%$, indicating that in the pristine sample the BP is less pronounced, blue shifted and broader.
The observed dependence on stability follows the same trend reported for hyperquenched glasses \cite{DOShyper} and simulated on physical vapor deposited (PVD) glasses \cite{singh_ultrastable_2013}.
The BP intensity depletion in pristine amber is at ease with the 22$\%$ decrease of the low temperature bump in
$C_p/T^3$ \cite{perez}, and it can not be accounted for by a simple Debye $\omega_D^{-3}$ scaling (which would imply only a 7.4$\%$ depletion).
Previous works have reported BP blueshift and intensity depletion after strong pressure-induced densification \cite{BPGeO2dense,Liu1995}. However, contrary to what we observe, the density increase 
doesn't necessarily imply the thermodynamical stabilization testified, here, by the $T_f$ reduction of pristine amber.

Large experimental and theoretical efforts have been devoted to the study of the connection of the BP with the elastic properties of disordered media (although the BP has been also related to the Van Hove singularities of crystals \cite{PhysRevLett.106.225501}). For instance, the BP position has been correlated to the longitudinal \cite{ruffle_glass-specific_2006} and transverse \cite{shintani_universal_2008} Ioffe-Regel limit, i.e. the condition met when the mean free path of the corresponding sound waves equal the size of their characteristic wavelengths. In the transverse case, numerical simulations in 2D glass-forming systems suggested that the BP height should scale with the inverse of the shear modulus. This is not the case for amber, where based on the available data on the sound velocity \cite{perez2016}, we would expect a BP reduction $< 7\%$, well below the experimentally observed 22$\%$ (Cp)\cite{perez} or 26$\%$ (IXS-NRA). The Soft-Potential Model (SPM) postulates the coexistence of acoustic phonons and quasi-localized modes, the latter being either low-energy tunneling two-level systems or soft vibrational modes with $g(E) \propto E^4$ below the BP energy\cite{Karpov1983,SPM}. Extensions of the SPM\cite{Parshin2007,Schober2014} attribute the ubiquitous appearance of a maximum in the $\frac{g(E)}{E^2}$ or BP, to the interaction among these quasi-localized modes themselves and the acoustic modes. The SPM framework has been successfully used to rationalize the low-temperature glassy behavior \emph{below} the BP temperature, but it does not provide any quantitative prediction about the BP temperature or frequency as to be checked in the present study. For example, the SPM predicts that the acoustic attenuation --like $g(E)$-- scales with $E^4$ due to the resonant absorption of sound waves by those soft vibrations, but only at energies below the BP. 

A theoretical model that investigates the frequency regime accessible via IXS experiments,
is the fluctuating elasticity theory (FET) that accounts for the disorder in terms of microscopic random spatial fluctuations of the transverse elastic constant (shear modulus) of the glass, with a certain degree of spatial correlation \cite{schirmacher2008vibrational}. Importantly, FET quantitatively relates the BP position to the broadness and correlation length of the elastic constant distribution of the disordered matrix, factoring out any density dependence. Hence, our observation of vibrational properties (v-DOS, sound attenuation, sound dispersion) dependence on fictive temperature, combined with the FET predictions, allows for ultimately establishing a connection between stability amount and correlation of dynamical disorder of the glass.
FET can be solved for a gaussian distribution of elastic constants with variance $\gamma$, correlated within a characteristic lengthscale $\xi$. The resulting BP, evaluated as detailed in SI, is reported in the inset of Fig.\ref{f:DOS}. Best agreement with experimental data is obtained for a disorder reduction from $\gamma_{an}=0.477 \pm 0.005$ to $\gamma_{pr}=0.451 \pm 0.007$ and lengthscale increase from $\xi_{an}=1.7 \pm 0.1$ nm to $\xi_{pr}=2.1 \pm 0.1$nm going from the annealed to the pristine sample. The blueshifted and less pronounced BP of pristine amber, indicates therefore that vitrification obtained in lower regions of the energy landscape (i.e. lower fictive temperatures) corresponds to a sharper and longer-range correlated distribution of elastic constants (a less disordered structure). 

Importantly, the comparison of acoustic properties of pristine and annealed amber provides a further benchmark for the correlation between fictive temperature and dynamical disorder. FET predicts a tight correlation between the BP and hypersonic attenuation in the THz regime for a given disorder distribution \cite{schirmacher_acoustic_2007}, in line with different theoretical approaches such as the SPM \cite{PhysRevLett.100.015501}, and with 
experimental studies on glasses with different fragility values, also indicating a correlation of the BP strength with acoustic damping \cite{doi:10.1063/1.2912060}. Accordingly, we investigated collective excitations in the high frequency (THz) regime by means of IXS at the beamline ID28 of ESRF, determining the sound dispersion and attenuation of the longitudinal density fluctuations. 

A selection of the inelastic x-ray scattering spectra of pristine amber for constant values of the exchanged momentum, Q, is shown in SI.
\begin{figure}[ht]
	\begin{center}
		\includegraphics[width=0.48\textwidth]{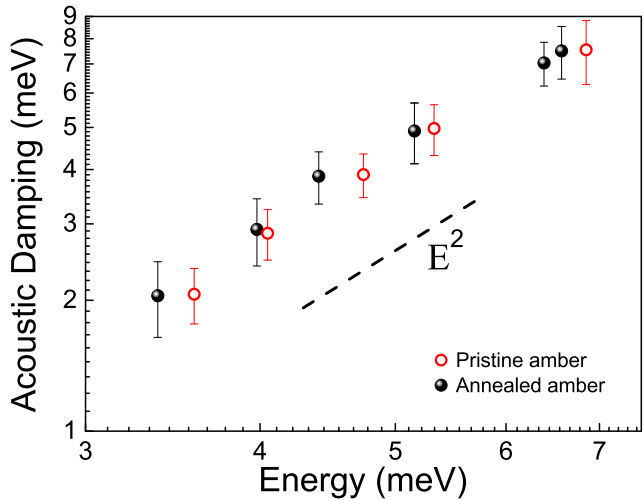}
		\caption{The acoustic damping $\Gamma(E)$ in the IXS frequency range at room temperature for pristine (red open dots) and annealed (black dots) amber. Dashed line is a power-2 steepness as a guide for the eyes.}
		\label{f:attenuation}
	\end{center}
\end{figure}
The linewidth $\Gamma(Q)$ of the inelastic peaks, corresponding to the acoustic attenuation, is reported in Fig.\ref{f:attenuation}. In both pristine and annealed amber, $\Gamma(Q)$ exhibits a similar power law dependence on the acoustic energy $E(Q)$ compatible with a quadratic law. This behaviour is well documented in a number of glassy systems and it is ascribable to the dominant role of the structural disorder in the THz frequency regime \cite{PhysRevLett.83.5583,natcom}.

Critically, FET predicts that larger width of the elastic constants distribution result in larger attenuation. While the dependency is maximum for frequencies at the onset of the BP, and tends to vanish in the high frequency limit, the trend still holds on the blue side of the BP \cite{schirmacher_acoustic_2007,schirmacher2008vibrational}, i.e. the region probed by IXS. Hence, our experimental observation of a slightly lower hypersonic damping in pristine amber is a further important indication that lower fictive temperatures correspond to less disordered structures.

\begin{figure}[ht]
	\begin{center}
		\includegraphics[width=0.48\textwidth]{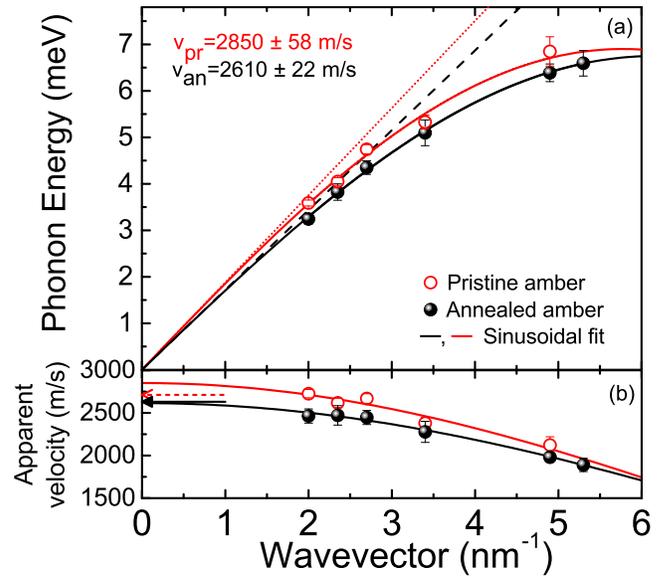}
		\caption{(a) Acoustic modes dispersion of pristine (red open dots) and annealed (black dots) amber. Error bars from the DHO fit are reported (see SI for further details). Solid lines are the sinusoidal function best fitted the modes' dispersions, dashed lines are the derivative at Q $\rightarrow 0$;(b) Apparent sound velocity $\frac{E}{Q}$ dispersion for pristine (open dots) and annealed (black dots) amber, together with the sinusoidal best fit functions of panel \textit{a} divided by Q (solid lines). The two arrows indicate the velocities measured by Brillouin light scattering (red dashed arrow $v_{pr}$=2710 m/s and black solid arrow $v_{an}=2630$ m/s) from Ref. \cite{perez2016}. The Q $\rightarrow 0$ limit shows positive dispersion for the pristine, not for the annealed amber.}
		\label{f:dispersion}
	\end{center}
\end{figure}

A third benchmark for the observed correlation between fictive temperature and elastic constants distribution is provided by the sound velocity behaviour reported in  Fig.\ref{f:dispersion}a.
The energy peak position $E(Q)$ of the inelastic peaks, of both pristine and annealed amber, exhibits a dispersion which testifies the propagating nature of the acoustic modes. The dispersion bends down reaching a maximum for Q around half of Q$_{p}$ (5.8 nm$^{-1}$ for the pristine, 6.1 nm$^{-1}$ for the annealed amber), identifying a pseudo-Brillouin zone.
The apparent velocity $E(Q)/Q$ is reported in Fig.\ref{f:dispersion}b, along with the low-frequency sound velocities (v$_{pr}=2710$ m/s and v$_{an}=2630$ m/s) measured by Brillouin light scattering \cite{perez2016}. The observed trend is in line with the recently reported experimental correlation between sound velocity and $T_f$ \cite{Pogna2,kearns2010}. The longitudintal sound velocity value of pristine amber in the IXS
characteristic Q-region, $c_{\infty}$, exceeds the value expected in
the $Q \rightarrow 0$ limit, $c_{0}$, by nearly 5\%.
This \textit{positive dispersion} has been explained as a signature of a relaxation processes related to the presence of microscopic disorder, reported also in purely harmonic model glasses \cite{goetzePRE}. Elastic constant disorder, indeed, is the leading mechanism of high frequency sound attenuation in glasses and, in turn, it is the manifestation of relaxations by the spontaneous density fluctuations occurring over some characteristic timescale $\tau$. When the frequency of the acoustic wave $\omega (Q)=E/\hbar$ crosses the value $\tau^{-1}$, the sound velocity increases by a few percents \cite{Scopigno2002341,goetzePRE} with respect to the relaxed value $c_{0}$.
The mismatch between the Q$\rightarrow 0$ extrapolation of the dispersion relation and the measured low-frequency sound velocity indicates the presence of positive dispersion for the pristine amber. This has to be contrasted with the behaviour of the annealed glass, showing -over the same Q window- the same sound velocity of its relaxed limit ($Q \rightarrow 0$). Critically, this observation implies that, in our experiment, the condition $\tau_{an} < \frac{1}{\omega (Q) } < \tau_{pr}$ holds. The larger relaxation time of the pristine sample indicates longer living acoustic waves compared to the annealed sample; a further indication of less disordered distribution of elastic constants upon stabilization.
\begin{figure}[ht]
	\begin{center}
		\includegraphics[width=0.48\textwidth]{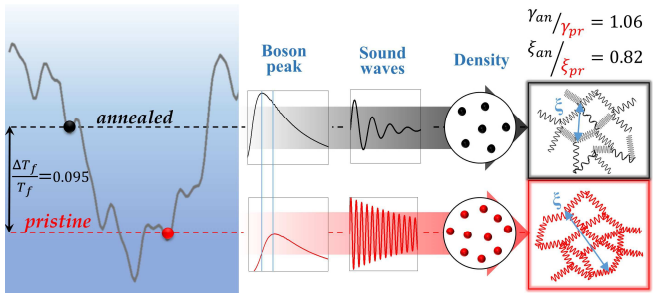}
		\caption{Summary of structural and dynamical properties as function of the position in the landscape which, in turn, relates to $T_f$. Ageing-induced stabilization results in a density increase, while Boson peak and sound attenuation are both reduced. Correspondingly, the distribution of elastic disorder gets narrower and longer range correlated.}
		\label{f:conclusion}
	\end{center}
\end{figure}
In conclusion, we experimentally assessed how the vibrational properties of a glass undergo a sizable modification via natural hyperaging. A significant redistribution of the vibrational density of states, testified by a depletion, blue-shift and broadening of the BP, accompanies the homogeneous densification occurring during amber's stabilization. The reported effects are different in nature from the BP alterations induced by hydrostatic compression. The changes of the vibrational density of states, here, are associated with a modification of the acoustic excitations, which we ultimately trace back to the dependence of the elastic disorder (width and correlation length of its distribution) on the energy landscape level. Taken all together (Fig.\ref{f:conclusion}), our findings reveal that a 9\% lowering of the fictive temperature upon stabilization correspond to a 6\% ordering of the elastic matrix and a 22\% increase of its fluctuations lengthscale.

\section{acknowledgement}
	The authors are grateful to G. Parisi and to R.J. Jim\'enez-Riob\'oo for helpful discussions, to J.M. Castilla for his help preparing the amber samples, and to Alfredo Argumosa (El Soplao, S.L., Santander) for providing  amber samples. The authors acknowledge A. Bosak for the support to the inelastic x-ray scattering measurements.
	M.A.R. acknowledges financial support from the Spanish Ministry of Economy and Competitiveness through projects FIS2017-84330-R and "Mar\'ia de Maeztu" Programme for Units of Excellence in R\&D (MDM-2014-0377), as well as from the Autonomous Community of Madrid through programme NANOFRONTMAG-CM (S2013/MIT-2850). This project has received funding from the European Union's Horizon 2020 research and innovation programme under grant agreement No. 785219 - GrapheneCore2

\newpage
\section{Supplementary Information}
\subsection*{Sample preparation and annealing}
The amber samples measured typically consist of $\sim 10 \times 10 \times 2 mm^3$ plates (masses $\sim$ 0.1--0.3 g) whose lateral dimension matches the absorption length at $\sim$23 keV, i.e. providing the ideal sample to maximize the IXS signal. The pristine amber comes from the deposit of El Soplao, in Cantabria Spain \cite{ambra_quiS,ElSoplaoS} dated to be 110-112 million years old and it corresponds to resin produced by the extinct Cheirolepidiaceae (amber type B). The annealing of the pristine sample is achieved by keeping the amber at $T_g$ + 10K (433K) for 3 hours and by cooling it to room temperature at the standard rate of 10 K/min. The thermal treatment is conducted in vacuum-sealed pyrex ampoules in a 400 mbar N$_2$ gas atmosphere.
\subsection*{DSC scan and fictive temperature determination}
Calorimetric characterization of amber was performed using DSC upscans.
Heating and cooling rates employed were always $\pm$ 10 K/min.
The accuracy of temperature control is estimated to be better than 0.1 K.
%
The fictive temperature, $T_f$, is defined as the temperature at which the non-equilibrium (glass) state and its equilibrium (supercooled liquid) state would have the same structure and properties, in particular the same enthalpy \cite{tulS,moyS,Guo20113230S}.

\begin{figure}[htp!]
	\includegraphics[width=0.5\textwidth]{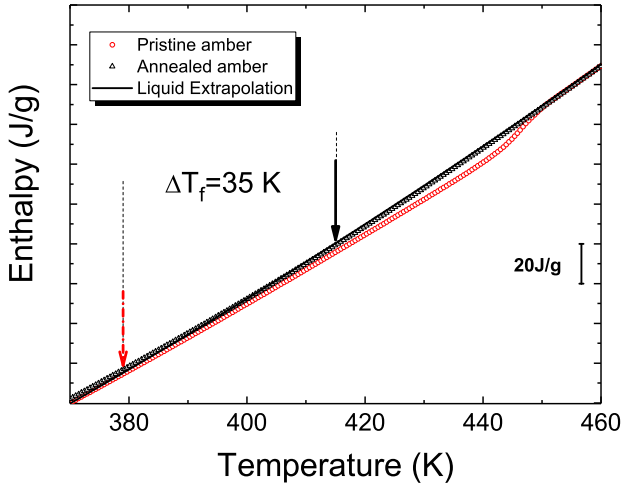}
	\caption{Fictive temperature determination of pristine (red dashed arrow) and annealed (black arrow) amber, following ref.\cite{perezS} , from the intersection of the glass and liquid enthalpy curves.}
	\label{fig:H}
\end{figure}

Accordingly, $T_f$'s of pristine and annealed amber can be obtained from the enthalpy curves obtained by direct integration of the specific-heat curves.
The starting point for integration is T = 465 K, well above $T_g$ = 423 K, and the aging signal peaked at $T_{max}$ = 437 K.
The fictive temperature is obtained from the intersection of the extrapolations of the liquid and the glass enthalpy curves, see Fig. S\ref{fig:H}. The calculation of the extrapolations was done using experimental data far from the glass transition, by means of quadratic polynomial fits, given the linear behavior of the specific-heat curves well in the liquid and glass regions. Specifically, we used the temperature ranges 458 K $\leq$T$\leq$ 465 K and 320 K$\leq$ T$\leq$ 355 K for the liquid and glass extrapolations. In Fig. S\ref{fig:H} we can observe that enthalpy curves for all samples regardless of their thermal history collapse for temperatures above 450 K. The errors involved in the determination of $T_f$ are estimated to be 1 K due to statistical errors in the extrapolations of glass and liquid enthalpy curves. The lower $T_f$ of pristine amber proves for the higher thermodynamical stability.

\subsection*{Inelastic x-ray scattering with nuclear resonance analysis}
The IXS-NRA experiments were performed at the beam line ID18 of ESRF with the storage ring working in 16-bunch mode. A description of the  technique can be found in ref. \cite{rufleINSS}. 

\begin{figure}[h!]
	\begin{center}
		\includegraphics[width=0.5\textwidth]{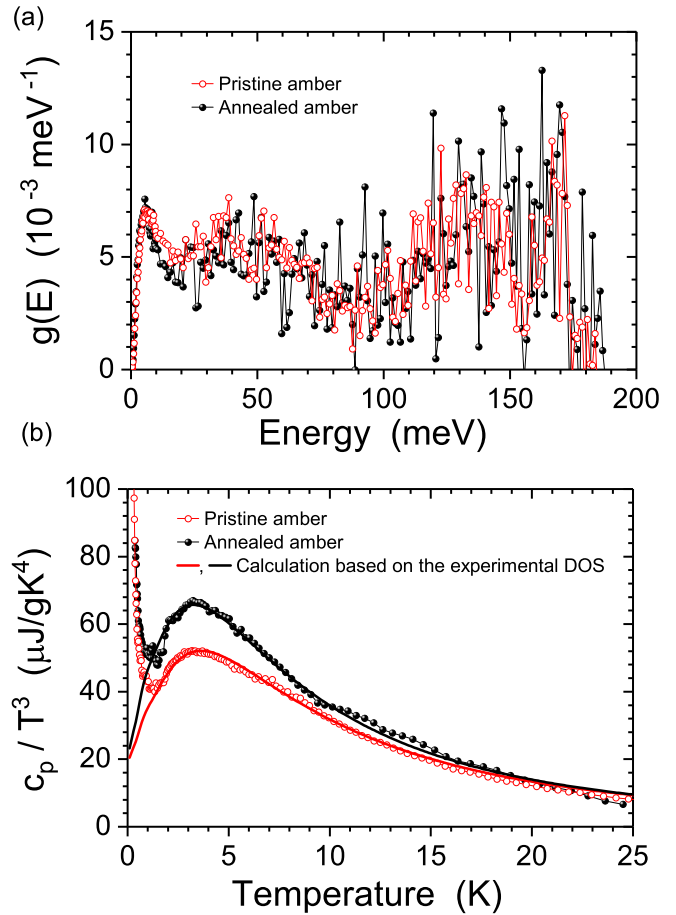}
		\caption{(a) Vibrational density of states of annealed (black dots) and pristine amber (red open dots); (b) Experimental specific heat of pristine (red open dots) and annealed amber (black dots) compared to the calculated value from the VDOS (solid lines).}
		\label{fig:cp}
	\end{center}
\end{figure}
The energy spectra of inelastic x-ray scattering were collected over a range of -10 to 190 meV around the $^{57}$Fe nuclear resonance energy of 14.4 keV with the energy resolution of 0.70 meV. From the measured spectra, the density of vibrational states was derived using a standard double-Fourier transformation procedure \cite{Kohn2000S} using as single variable parameter an effective recoil energy.  Initially this parameter was estimated from the mean atomic mass of amber and the mean scattering angle, and the final value (19.6 meV) was determined from the best fit of the specific heat calculated from the VDOS to the experimental value. Non reduced vibrational density of states and the fit of the experimental heat capacity to calculations based on it are presented in Fig. S\ref{fig:cp}. Measurements were performed in vacuum at a constant temperature of T=30 K with a total integration time of about 800 min per sample.
\begin{figure}[h!]
	\begin{center}
		\includegraphics[width=0.5\textwidth]{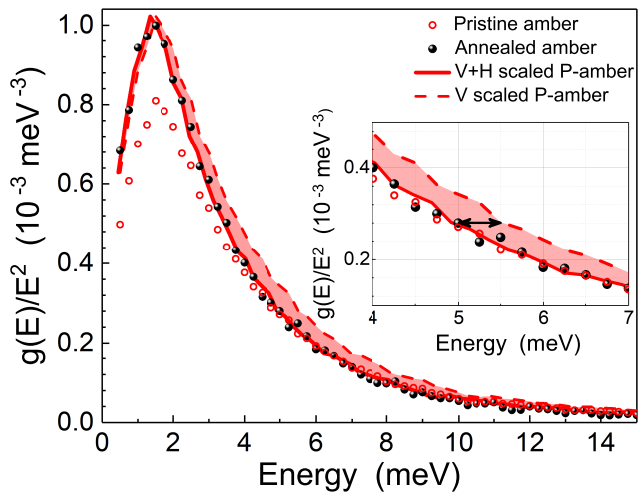}
		\caption{Reduced DOS (BP)of pristine (red open dots) and annealed
			amber (black dots) at 30K measured by IXS-NRA. A vertical scaling of pristine data (expansion of 26\% reported with red dashed line) is not sufficient to overlap the two BP. As emphasized in the inset, a shrinking of the pr-amber energy scale is also necessary. The experimental data of pristine scaled vertically and horizontally are reported with red solid line.}
		\label{fig:scaling}
	\end{center}
\end{figure}
The experimental data of pristine and annealed amber can be overlapped by relative scaling of both vertical and horizontal axes, as shown in Fig. S\ref{fig:scaling}.

\subsection*{Inelastic x-ray scattering}
\begin{figure}[ht]
	\begin{center}
		\includegraphics[width=0.48\textwidth]{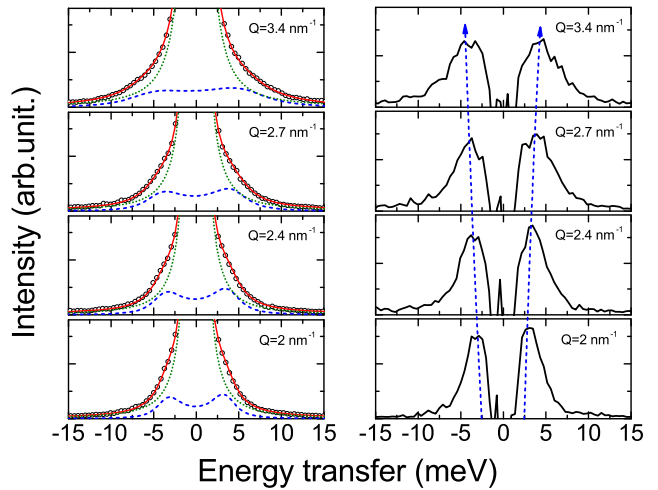}
		\caption{IXS spectra at T=295 K for selected momentum transfer Q. Left: data points (open dots) are fitted with damped harmonic oscillator function (DHO-blue dashed line) added to a delta function (green dot line) to account for inelastic mode and elastic scattering, respectively, all convoluted with instrumental resolution. Acoustic mode energy and attenuation are obtained from the DHO parameters. The DHO energy-parameter, corresponds to the position of current's $\omega^2 S(\omega,Q)$ maxima. Right: longitudinal excitations (black line), obtained by the raw IXS spectra after elastic component subtraction. Blue dashed arrows are added to highlight the peak position change with Q.}
		\label{amber_spectra}
	\end{center}
\end{figure}

The IXS experiments were performed at the beam line ID28 of ESRF. The frequency spectrum of the scattered x-ray intensity which is  proportional to the dynamic structure factor $S(Q,\hbar \omega$), is measured as function of the exchanged momentum $Q$ defined by the scattering angle $\theta$ and by the wavevector of the incident photons $k_i$ as $Q=2k_i sin (\theta /2)$. The beam was focused down to ~$50 \times 250\mu$m$^2$ and a 8-analysers bench was used.
The scattering intensity I(Q) shown in Fig. S\ref{amber_spectra} is measured by scanning the scattering angle at constant energy $E=0$ and it is proportional to the static structure factor S(Q) through the form factor f(Q). Different absolute intensity of I(Q) of pristine and annealed amber can follow from different sample position and small deviation of from the elastic line of the detectors in the two measurements. 
Energy scans were performed at constant $Q$ values in the range $1$ to 7 $\approx$ nm$^{-1}$, mapping more than half of the pseudo Brillouin Zone (Q$\sim 10$ nm$^{-1}$). The momentum resolution of $0.25$ nm$^{-1}$ was determined by slits placed in front of the analyzer.
The scanned energy range was $-30~\leq \hbar \omega\leq~30$ meV, where $\hbar \omega=E_0 - E$ is the energy transfer, with $E_0$ and $E$ being the energy of the incident ($23.725$ eV) and the scattered x-ray photons. Using the (12 12 12) reflection for the Si monochromator and crystal analyzers the overall energy resolution was $1.5$ meV, corresponding to the FWHM of instrumental function (green dot line under the spectra in Fig. S\ref{amber_spectra}). Measurements were performed at room temperature ($T=~295$ K).

The spectral intensity, $S(Q,\omega)$, is dominated by an intense elastic peak on top of an inelastic Brillouin component. The origin of the elastic and inelastic spectral components in the high frequency regime probed by IXS has been thoroughly discussed in the literature \cite{sette_dynamics_1998S}. In a nutshell, the scattering intensity is the Fourier transform of the density autocorrelator, $F(Q,t)$. The vibrational dynamics in the glassy state is responsible for an oscillatory component of $F(Q,t)$ on the picosecond timescale, which is damped in view of the non-plane wave nature of the vibrational eigenmodes due to the disorder. This oscillatory component corresponds to the broad Brillouin doublet observed in the IXS spectrum. On a longer timescale, $F(Q,t)$ reaches a plateau, the non ergodicity factor, representing the residual amount of decorrelation due to the structural arrest characterizing the glassy state. This plateau gives rise to the elastic component in the $S(Q,\omega)$ \cite{scopigno_fragilityS,scopigno_universal_2010S}. Accordingly, the spectra have been analysed with a damped harmonic oscillator model (DHO)\cite{scopigno_microscopic_2005S} in order to extract the vibrational modes' energy and attenuation jointly with a delta function to account for the elastic scattering. Both components were convoluted with the instrumental resolution function and corrected for the detailed balance condition. 

\section*{Scaling of the static structure factor}

\begin{figure}[!ht]
	\begin{center}
		\includegraphics[width=0.48\textwidth]{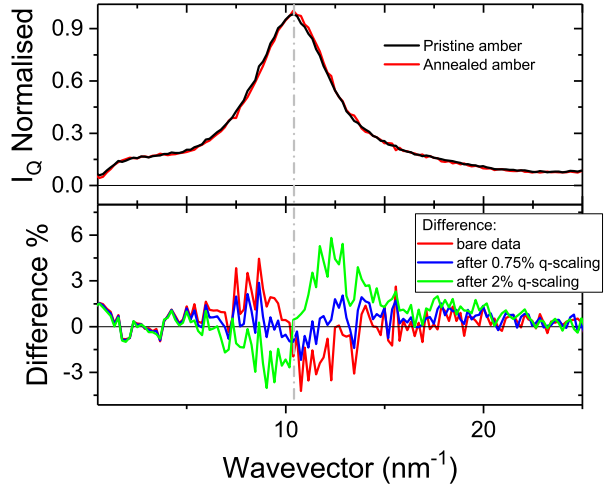}
		\caption{X-ray diffraction. Top: Data for pristine (red line) and corresponding annealed amber (black line) are reported. Bottom: Relative differences for various x-axis scaling factors are shown.}
		\label{s(q)scaling}
	\end{center}
\end{figure}

Diffraction data have been first intensity normalized to the main peak.  
The densification caused by hydrostatic compression is often achieved by changes of the intermediate range order signaled by a shift of the first diffraction peak, linearly dependent on the density (for example in tetrahedral glasses). In order to verify whether such linearity also applies in the case of the 2\% densification of the pristine material achieved by natural ageing, we applied a 2\% relative q-scaling. Clearly, this is not the case, as the pristine-annealed difference is minimized with a relative (best fitted) 0.75\% x-axis scaling as shown in Fig. \ref{s(q)scaling}. 
\subsection*{Fluctuating elasticity theory}
The fluctuating elasticity theory (FET) describes a glass in terms of an elastic matrix characterized by random spatial fluctuation of the elastic constants $v^2(r)$ ($v(r)$ is a sound velocity), $\Delta(r)$, which are spatially correlated over a lenghtscale $\xi$:
\begin{subequations}
	\begin{align} 
	v^2(r)&=v^2_0+\Delta(r)\\
	\langle\Delta(r+r_0)\Delta(r_0)\rangle&=C(r)=\langle\Delta^2\rangle e^{-r/\xi}
	\end{align} 
\end{subequations}
Within FET, the vibrational density of states reads:
\begin{equation}
g(\omega)=\frac{2\omega}{\pi}\mathrm{Im}\left[\int\left(\frac{dk}{2\pi}\right)^3\frac{1}{-z^2+k^2(v_0^2-\Sigma(z))}\right]
\label{DOS} 
\end{equation} 
$\Sigma(z)$ is the low $q$ limit of the so called self energy, it renormalizes $v_0$ to match the experimental sound velocity $v_{exp}^2=v_0^2-\mathrm{Re}(\Sigma(z\approx 0))$, and can be evaluated in the self consistent Born approximation (SCBA) as:
\begin{equation}
\Sigma(z)=\frac{\gamma}{2}\frac{v_0^4}{\varphi_d}\int\displaylimits_{|\mathbf{\mathrm{k}}|<k_D}\left(\frac{dk}{2\pi}\right)^3\frac{k^2C(k)}{\langle\Delta^2\rangle}\frac{1}{-z^2+k^2(v_0^2-\Sigma(z))}
\end{equation}
where $k_D$ is the Debye wavevector, $\varphi_d=\int\displaylimits_{|\mathbf{\mathrm{k}}|<k_D}\left(\frac{dk}{2\pi}\right)^3\frac{C(k)}{\langle\Delta^2\rangle}$ and $\gamma={\langle\Delta^2\rangle}\varphi_d/v_0^4$ is the disorder parameter.
The lineshape predicted by Eq.(\ref{DOS}) is fitted to the experimental v-DOS of pristine and annealed samples in  Fig. 2 of the main text, using the corresponding experimental densities, $\rho$, and Debye velocities \cite{perez2016S}. The free parameters are $\gamma$ and $\xi$ for each sample, and the debye wavevectors with a constrain on their ratio, imposed by their relative densities in view of the dependence $k_D\propto\rho^{1/3}$. The relative error on the estimation of $k_D$ extracted from the $95\%$ confidence intervals is $1.9\%$.


\begin{thebibliography}{51}
	\expandafter\ifx\csname natexlab\endcsname\relax\def\natexlab#1{#1}\fi
	\expandafter\ifx\csname bibnamefont\endcsname\relax
	\def\bibnamefont#1{#1}\fi
	\expandafter\ifx\csname bibfnamefont\endcsname\relax
	\def\bibfnamefont#1{#1}\fi
	\expandafter\ifx\csname citenamefont\endcsname\relax
	\def\citenamefont#1{#1}\fi
	\expandafter\ifx\csname url\endcsname\relax
	\def\url#1{\texttt{#1}}\fi
	\expandafter\ifx\csname urlprefix\endcsname\relax\def\urlprefix{URL }\fi
	\providecommand{\bibinfo}[2]{#2}
	\providecommand{\eprint}[2][]{\url{#2}}
	
	\bibitem[{\citenamefont{Goldstein}(1969)}]{doi:10.1063/1.1672587}
	\bibinfo{author}{\bibfnamefont{M.}~\bibnamefont{Goldstein}},
	\bibinfo{journal}{J. Chem. Phys.} \textbf{\bibinfo{volume}{51}},
	\bibinfo{pages}{3728} (\bibinfo{year}{1969}).
	
	\bibitem[{\citenamefont{Stillinger}(1995)}]{Stillinger1935}
	\bibinfo{author}{\bibfnamefont{F.~H.} \bibnamefont{Stillinger}},
	\bibinfo{journal}{Science} \textbf{\bibinfo{volume}{267}},
	\bibinfo{pages}{1935} (\bibinfo{year}{1995}).
	
	\bibitem[{\citenamefont{Corezzi et~al.}(2002)\citenamefont{Corezzi, Fioretto,
			and Rolla}}]{nature_vitri}
	\bibinfo{author}{\bibfnamefont{S.}~\bibnamefont{Corezzi}},
	\bibinfo{author}{\bibfnamefont{D.}~\bibnamefont{Fioretto}}, \bibnamefont{and}
	\bibinfo{author}{\bibfnamefont{P.}~\bibnamefont{Rolla}},
	\bibinfo{journal}{Nature} \textbf{\bibinfo{volume}{420}},
	\bibinfo{pages}{653} (\bibinfo{year}{2002}).
	
	\bibitem[{\citenamefont{Zhao et~al.}(2013{\natexlab{a}})\citenamefont{Zhao,
			Ragazzi, and McKenna}}]{Zhao20137041}
	\bibinfo{author}{\bibfnamefont{J.}~\bibnamefont{Zhao}},
	\bibinfo{author}{\bibfnamefont{E.}~\bibnamefont{Ragazzi}}, \bibnamefont{and}
	\bibinfo{author}{\bibfnamefont{G.~B.} \bibnamefont{McKenna}},
	\bibinfo{journal}{Polymer} \textbf{\bibinfo{volume}{54}},
	\bibinfo{pages}{7041} (\bibinfo{year}{2013}{\natexlab{a}}).
	
	\bibitem[{\citenamefont{Zhao et~al.}(2013{\natexlab{b}})\citenamefont{Zhao,
			Simon, and McKenna}}]{Mckenna}
	\bibinfo{author}{\bibfnamefont{J.}~\bibnamefont{Zhao}},
	\bibinfo{author}{\bibfnamefont{S.~L.} \bibnamefont{Simon}}, \bibnamefont{and}
	\bibinfo{author}{\bibfnamefont{G.~B.} \bibnamefont{McKenna}},
	\bibinfo{journal}{Nat. Commun.} \textbf{\bibinfo{volume}{4}},
	\bibinfo{pages}{1783} (\bibinfo{year}{2013}{\natexlab{b}}).
	
	\bibitem[{\citenamefont{P\'erez-Casta$\tilde{n}$eda
			et~al.}(2013)\citenamefont{P\'erez-Casta$\tilde{n}$eda, Jim\'enez-Riob\'oo,
			and Ramos}}]{ramos_velocity}
	\bibinfo{author}{\bibfnamefont{T.}~\bibnamefont{P\'erez-Casta$\tilde{n}$eda}},
	\bibinfo{author}{\bibfnamefont{R.~J.} \bibnamefont{Jim\'enez-Riob\'oo}},
	\bibnamefont{and} \bibinfo{author}{\bibfnamefont{M.~A.} \bibnamefont{Ramos}},
	\bibinfo{journal}{J. Phys.: Condens. Matter} \textbf{\bibinfo{volume}{25}},
	\bibinfo{pages}{295402} (\bibinfo{year}{2013}).
	
	\bibitem[{\citenamefont{Zeller and Pohl}(1971)}]{zeller_thermal_1971-1}
	\bibinfo{author}{\bibfnamefont{R.~C.} \bibnamefont{Zeller}} \bibnamefont{and}
	\bibinfo{author}{\bibfnamefont{R.~O.} \bibnamefont{Pohl}},
	\bibinfo{journal}{Phys. Rev. B} \textbf{\bibinfo{volume}{4}},
	\bibinfo{pages}{2029} (\bibinfo{year}{1971}).
	
	\bibitem[{\citenamefont{Binder and Kob}(2005)}]{binder_glassy_2005}
	\bibinfo{author}{\bibfnamefont{K.}~\bibnamefont{Binder}} \bibnamefont{and}
	\bibinfo{author}{\bibfnamefont{W.}~\bibnamefont{Kob}},
	\emph{\bibinfo{title}{Glassy Materials And Disordered Solids: An Introduction
			to Their Statistical Mechanics}} (\bibinfo{publisher}{World Scientific
		Publishing Co.: Singapore}, \bibinfo{year}{2005}), ISBN
	\bibinfo{isbn}{9812565108}.
	
	\bibitem[{\citenamefont{J{\"a}ckle}(1981)}]{amorph_sol}
	\bibinfo{author}{\bibfnamefont{J.}~\bibnamefont{J{\"a}ckle}},
	\emph{\bibinfo{title}{Low Frequency Raman Scattering in Glasses}}
	(\bibinfo{publisher}{Springer Verlag: Heidelberg, Germany},
	\bibinfo{year}{1981}).
	
	\bibitem[{\citenamefont{Malinovsky and Sokolov}(1986)}]{malinovsky1986nature}
	\bibinfo{author}{\bibfnamefont{V.~K.} \bibnamefont{Malinovsky}}
	\bibnamefont{and} \bibinfo{author}{\bibfnamefont{A.~P.}
		\bibnamefont{Sokolov}}, \bibinfo{journal}{Solid State Commun.}
	\textbf{\bibinfo{volume}{57}}, \bibinfo{pages}{757} (\bibinfo{year}{1986}).
	
	\bibitem[{\citenamefont{Surovtsev et~al.}(2003)\citenamefont{Surovtsev,
			Shebanin, and Ramos}}]{PhysRevB.67.024203}
	\bibinfo{author}{\bibfnamefont{N.~V.} \bibnamefont{Surovtsev}},
	\bibinfo{author}{\bibfnamefont{A.~P.} \bibnamefont{Shebanin}},
	\bibnamefont{and} \bibinfo{author}{\bibfnamefont{M.~A.} \bibnamefont{Ramos}},
	\bibinfo{journal}{Phys. Rev. B} \textbf{\bibinfo{volume}{67}},
	\bibinfo{pages}{024203} (\bibinfo{year}{2003}).
	
	\bibitem[{\citenamefont{Carini~Jr et~al.}(2013)\citenamefont{Carini~Jr, Carini,
			D'Angelo, Tripodo, Di~Marco, Vasi, and Gilioli}}]{PhysRevLett.111.245502}
	\bibinfo{author}{\bibfnamefont{G.}~\bibnamefont{Carini~Jr}},
	\bibinfo{author}{\bibfnamefont{G.}~\bibnamefont{Carini}},
	\bibinfo{author}{\bibfnamefont{G.}~\bibnamefont{D'Angelo}},
	\bibinfo{author}{\bibfnamefont{G.}~\bibnamefont{Tripodo}},
	\bibinfo{author}{\bibfnamefont{G.}~\bibnamefont{Di~Marco}},
	\bibinfo{author}{\bibfnamefont{C.}~\bibnamefont{Vasi}}, \bibnamefont{and}
	\bibinfo{author}{\bibfnamefont{E.}~\bibnamefont{Gilioli}},
	\bibinfo{journal}{Phys. Rev. Lett.} \textbf{\bibinfo{volume}{111}},
	\bibinfo{pages}{245502} (\bibinfo{year}{2013}).
	
	\bibitem[{\citenamefont{Buchenau et~al.}(1984)\citenamefont{Buchenau, N\"ucker,
			and Dianoux}}]{PhysRevLett.53.2316}
	\bibinfo{author}{\bibfnamefont{U.}~\bibnamefont{Buchenau}},
	\bibinfo{author}{\bibfnamefont{N.}~\bibnamefont{N\"ucker}}, \bibnamefont{and}
	\bibinfo{author}{\bibfnamefont{A.~J.} \bibnamefont{Dianoux}},
	\bibinfo{journal}{Phys. Rev. Lett.} \textbf{\bibinfo{volume}{53}},
	\bibinfo{pages}{2316} (\bibinfo{year}{1984}).
	
	\bibitem[{\citenamefont{Zorn et~al.}(1995)\citenamefont{Zorn, Arbe, Colmenero,
			Frick, Richter, and Buchenau}}]{zorn1995neutron}
	\bibinfo{author}{\bibfnamefont{R.}~\bibnamefont{Zorn}},
	\bibinfo{author}{\bibfnamefont{A.}~\bibnamefont{Arbe}},
	\bibinfo{author}{\bibfnamefont{J.}~\bibnamefont{Colmenero}},
	\bibinfo{author}{\bibfnamefont{B.}~\bibnamefont{Frick}},
	\bibinfo{author}{\bibfnamefont{D.}~\bibnamefont{Richter}}, \bibnamefont{and}
	\bibinfo{author}{\bibfnamefont{U.}~\bibnamefont{Buchenau}},
	\bibinfo{journal}{Phys. Rev. E} \textbf{\bibinfo{volume}{52}},
	\bibinfo{pages}{781} (\bibinfo{year}{1995}).
	
	\bibitem[{\citenamefont{Ramos et~al.}(1997)\citenamefont{Ramos, Vieira,
			Bermejo, Dawidowski, Fischer, Schober, Gonz\'alez, Loong, and
			Price}}]{ramos_cp}
	\bibinfo{author}{\bibfnamefont{M.~A.} \bibnamefont{Ramos}},
	\bibinfo{author}{\bibfnamefont{S.}~\bibnamefont{Vieira}},
	\bibinfo{author}{\bibfnamefont{F.~J.} \bibnamefont{Bermejo}},
	\bibinfo{author}{\bibfnamefont{J.}~\bibnamefont{Dawidowski}},
	\bibinfo{author}{\bibfnamefont{H.~E.} \bibnamefont{Fischer}},
	\bibinfo{author}{\bibfnamefont{H.}~\bibnamefont{Schober}},
	\bibinfo{author}{\bibfnamefont{M.~A.} \bibnamefont{Gonz\'alez}},
	\bibinfo{author}{\bibfnamefont{C.~K.} \bibnamefont{Loong}}, \bibnamefont{and}
	\bibinfo{author}{\bibfnamefont{D.~L.} \bibnamefont{Price}},
	\bibinfo{journal}{Phys. Rev. Lett.} \textbf{\bibinfo{volume}{78}},
	\bibinfo{pages}{82} (\bibinfo{year}{1997}).
	
	\bibitem[{\citenamefont{Scopigno et~al.}(2011)\citenamefont{Scopigno, Steurer,
			Yannopoulos, Chrissanthopoulos, Krisch, Ruocco, and
			Wagner}}]{scopigno2011vibrational}
	\bibinfo{author}{\bibfnamefont{T.}~\bibnamefont{Scopigno}},
	\bibinfo{author}{\bibfnamefont{W.}~\bibnamefont{Steurer}},
	\bibinfo{author}{\bibfnamefont{S.~N.} \bibnamefont{Yannopoulos}},
	\bibinfo{author}{\bibfnamefont{A.}~\bibnamefont{Chrissanthopoulos}},
	\bibinfo{author}{\bibfnamefont{M.}~\bibnamefont{Krisch}},
	\bibinfo{author}{\bibfnamefont{G.}~\bibnamefont{Ruocco}}, \bibnamefont{and}
	\bibinfo{author}{\bibfnamefont{T.}~\bibnamefont{Wagner}},
	\bibinfo{journal}{Nat. Commun.} \textbf{\bibinfo{volume}{2}},
	\bibinfo{pages}{195} (\bibinfo{year}{2011}).
	
	\bibitem[{\citenamefont{Chumakov et~al.}(2011)\citenamefont{Chumakov, Monaco,
			Monaco, Crichton, Bosak, R\"uffer, Meyer, Kargl, Comez, Fioretto
			et~al.}}]{PhysRevLett.106.225501}
	\bibinfo{author}{\bibfnamefont{A.~I.} \bibnamefont{Chumakov}},
	\bibinfo{author}{\bibfnamefont{G.}~\bibnamefont{Monaco}},
	\bibinfo{author}{\bibfnamefont{A.}~\bibnamefont{Monaco}},
	\bibinfo{author}{\bibfnamefont{W.~A.} \bibnamefont{Crichton}},
	\bibinfo{author}{\bibfnamefont{A.}~\bibnamefont{Bosak}},
	\bibinfo{author}{\bibfnamefont{R.}~\bibnamefont{R\"uffer}},
	\bibinfo{author}{\bibfnamefont{A.}~\bibnamefont{Meyer}},
	\bibinfo{author}{\bibfnamefont{F.}~\bibnamefont{Kargl}},
	\bibinfo{author}{\bibfnamefont{L.}~\bibnamefont{Comez}},
	\bibinfo{author}{\bibfnamefont{D.}~\bibnamefont{Fioretto}},
	\bibnamefont{et~al.}, \bibinfo{journal}{Phys. Rev. Lett.}
	\textbf{\bibinfo{volume}{106}}, \bibinfo{pages}{225501}
	(\bibinfo{year}{2011}).
	
	\bibitem[{\citenamefont{L\'eonforte et~al.}(2006)\citenamefont{L\'eonforte,
			Tanguy, Wittmer, and Barrat}}]{PhysRevLett.97.055501}
	\bibinfo{author}{\bibfnamefont{F.}~\bibnamefont{L\'eonforte}},
	\bibinfo{author}{\bibfnamefont{A.}~\bibnamefont{Tanguy}},
	\bibinfo{author}{\bibfnamefont{J.~P.} \bibnamefont{Wittmer}},
	\bibnamefont{and} \bibinfo{author}{\bibfnamefont{J.-L.}
		\bibnamefont{Barrat}}, \bibinfo{journal}{Phys. Rev. Lett.}
	\textbf{\bibinfo{volume}{97}}, \bibinfo{pages}{055501}
	(\bibinfo{year}{2006}).
	
	\bibitem[{\citenamefont{Schirmacher et~al.}(2007)\citenamefont{Schirmacher,
			Ruocco, and Scopigno}}]{schirmacher_acoustic_2007}
	\bibinfo{author}{\bibfnamefont{W.}~\bibnamefont{Schirmacher}},
	\bibinfo{author}{\bibfnamefont{G.}~\bibnamefont{Ruocco}}, \bibnamefont{and}
	\bibinfo{author}{\bibfnamefont{T.}~\bibnamefont{Scopigno}},
	\bibinfo{journal}{Phys. Rev. Lett.} \textbf{\bibinfo{volume}{98}},
	\bibinfo{pages}{025501} (\bibinfo{year}{2007}).
	
	\bibitem[{\citenamefont{Shintani and Tanaka}(2008)}]{shintani_universal_2008}
	\bibinfo{author}{\bibfnamefont{H.}~\bibnamefont{Shintani}} \bibnamefont{and}
	\bibinfo{author}{\bibfnamefont{H.}~\bibnamefont{Tanaka}},
	\bibinfo{journal}{Nat. Mater.} \textbf{\bibinfo{volume}{7}},
	\bibinfo{pages}{870} (\bibinfo{year}{2008}).
	
	\bibitem[{\citenamefont{Marruzzo et~al.}(2013)\citenamefont{Marruzzo,
			Schirmacher, Fratalocchi, and Ruocco}}]{marruzzo_heterogeneous_2013}
	\bibinfo{author}{\bibfnamefont{A.}~\bibnamefont{Marruzzo}},
	\bibinfo{author}{\bibfnamefont{W.}~\bibnamefont{Schirmacher}},
	\bibinfo{author}{\bibfnamefont{A.}~\bibnamefont{Fratalocchi}},
	\bibnamefont{and} \bibinfo{author}{\bibfnamefont{G.}~\bibnamefont{Ruocco}},
	\bibinfo{journal}{Sci. Rep.} \textbf{\bibinfo{volume}{3}},
	\bibinfo{pages}{1407} (\bibinfo{year}{2013}), ISSN \bibinfo{issn}{2045-2322}.
	
	\bibitem[{\citenamefont{Navarro et~al.}(2010)\citenamefont{Navarro,
			Pe$\tilde{n}$alver, P\'erez-de La~fuente, Ortega-Blanco, Menor-Salv\'an,
			Barr\'on, Soriano, Rosales, L\'opez Del~Valle, Velasco et~al.}}]{ElSoplao}
	\bibinfo{author}{\bibfnamefont{M.}~\bibnamefont{Navarro}},
	\bibinfo{author}{\bibfnamefont{E.}~\bibnamefont{Pe$\tilde{n}$alver}},
	\bibinfo{author}{\bibfnamefont{R.}~\bibnamefont{P\'erez-de La~fuente}},
	\bibinfo{author}{\bibfnamefont{J.}~\bibnamefont{Ortega-Blanco}},
	\bibinfo{author}{\bibfnamefont{C.}~\bibnamefont{Menor-Salv\'an}},
	\bibinfo{author}{\bibfnamefont{E.}~\bibnamefont{Barr\'on}},
	\bibinfo{author}{\bibfnamefont{C.}~\bibnamefont{Soriano}},
	\bibinfo{author}{\bibfnamefont{I.}~\bibnamefont{Rosales}},
	\bibinfo{author}{\bibfnamefont{R.}~\bibnamefont{L\'opez Del~Valle}},
	\bibinfo{author}{\bibfnamefont{F.}~\bibnamefont{Velasco}},
	\bibnamefont{et~al.}, \bibinfo{journal}{Acta Geol. Sin. - Engl.}
	\textbf{\bibinfo{volume}{84}}, \bibinfo{pages}{959} (\bibinfo{year}{2010}).
	
	\bibitem[{\citenamefont{Menor-Salv\'an
			et~al.}(2010)\citenamefont{Menor-Salv\'an, Najarro, Velasco, Rosales, Tornos,
			and Simoneit}}]{ambra_qui}
	\bibinfo{author}{\bibfnamefont{C.}~\bibnamefont{Menor-Salv\'an}},
	\bibinfo{author}{\bibfnamefont{M.}~\bibnamefont{Najarro}},
	\bibinfo{author}{\bibfnamefont{F.}~\bibnamefont{Velasco}},
	\bibinfo{author}{\bibfnamefont{I.}~\bibnamefont{Rosales}},
	\bibinfo{author}{\bibfnamefont{F.}~\bibnamefont{Tornos}}, \bibnamefont{and}
	\bibinfo{author}{\bibfnamefont{B.~R.} \bibnamefont{Simoneit}},
	\bibinfo{journal}{Org. Geochem.} \textbf{\bibinfo{volume}{41}},
	\bibinfo{pages}{1089} (\bibinfo{year}{2010}).
	
	\bibitem[{\citenamefont{Sette et~al.}(1998)\citenamefont{Sette, Krisch,
			Masciovecchio, Ruocco, and Monaco}}]{sette_dynamics_1998}
	\bibinfo{author}{\bibfnamefont{F.}~\bibnamefont{Sette}},
	\bibinfo{author}{\bibfnamefont{M.~H.} \bibnamefont{Krisch}},
	\bibinfo{author}{\bibfnamefont{C.}~\bibnamefont{Masciovecchio}},
	\bibinfo{author}{\bibfnamefont{G.}~\bibnamefont{Ruocco}}, \bibnamefont{and}
	\bibinfo{author}{\bibfnamefont{G.}~\bibnamefont{Monaco}},
	\bibinfo{journal}{Science} \textbf{\bibinfo{volume}{280}},
	\bibinfo{pages}{1550} (\bibinfo{year}{1998}).
	
	\bibitem[{\citenamefont{Pogna et~al.}(2013)\citenamefont{Pogna,
			Rodr\'iguez-Tinoco, Rodr\'iguez-Viejo, and Scopigno}}]{pogna}
	\bibinfo{author}{\bibfnamefont{E.~A.~A.} \bibnamefont{Pogna}},
	\bibinfo{author}{\bibfnamefont{C.}~\bibnamefont{Rodr\'iguez-Tinoco}},
	\bibinfo{author}{\bibfnamefont{J.}~\bibnamefont{Rodr\'iguez-Viejo}},
	\bibnamefont{and} \bibinfo{author}{\bibfnamefont{T.}~\bibnamefont{Scopigno}},
	\bibinfo{journal}{Sci. Rep.} \textbf{\bibinfo{volume}{3}},
	\bibinfo{pages}{2518} (\bibinfo{year}{2013}).
	
	\bibitem[{\citenamefont{Chumakov et~al.}(1996)\citenamefont{Chumakov, Baron,
			R\"uffer, Gr\"unsteudel, Gr\"unsteudel, and Meyer}}]{rufleINS}
	\bibinfo{author}{\bibfnamefont{A.~I.} \bibnamefont{Chumakov}},
	\bibinfo{author}{\bibfnamefont{A.~Q.~R.} \bibnamefont{Baron}},
	\bibinfo{author}{\bibfnamefont{R.}~\bibnamefont{R\"uffer}},
	\bibinfo{author}{\bibfnamefont{H.}~\bibnamefont{Gr\"unsteudel}},
	\bibinfo{author}{\bibfnamefont{H.~F.} \bibnamefont{Gr\"unsteudel}},
	\bibnamefont{and} \bibinfo{author}{\bibfnamefont{A.}~\bibnamefont{Meyer}},
	\bibinfo{journal}{Phys. Rev. Lett.} \textbf{\bibinfo{volume}{76}},
	\bibinfo{pages}{4258} (\bibinfo{year}{1996}).
	
	\bibitem[{\citenamefont{Moynihan et~al.}(1976)\citenamefont{Moynihan, Easteal,
			De~Bolt, and Tucker}}]{moy}
	\bibinfo{author}{\bibfnamefont{C.~T.} \bibnamefont{Moynihan}},
	\bibinfo{author}{\bibfnamefont{A.~J.} \bibnamefont{Easteal}},
	\bibinfo{author}{\bibfnamefont{M.~A.} \bibnamefont{De~Bolt}},
	\bibnamefont{and} \bibinfo{author}{\bibfnamefont{J.}~\bibnamefont{Tucker}},
	\bibinfo{journal}{J. Am. Ceram. Soc.} \textbf{\bibinfo{volume}{59}},
	\bibinfo{pages}{12} (\bibinfo{year}{1976}).
	
	\bibitem[{\citenamefont{Perez-Casta$\tilde{n}$eda
			et~al.}(2014)\citenamefont{Perez-Casta$\tilde{n}$eda, Jimenez-Riob\'oo, and
			Ramos}}]{perez}
	\bibinfo{author}{\bibfnamefont{T.}~\bibnamefont{Perez-Casta$\tilde{n}$eda}},
	\bibinfo{author}{\bibfnamefont{R.~J.} \bibnamefont{Jimenez-Riob\'oo}},
	\bibnamefont{and} \bibinfo{author}{\bibfnamefont{M.~A.} \bibnamefont{Ramos}},
	\bibinfo{journal}{Phys. Rev. Lett.} \textbf{\bibinfo{volume}{112}},
	\bibinfo{pages}{165901} (\bibinfo{year}{2014}).
	
	\bibitem[{\citenamefont{Perez-Casta$\tilde{n}$eda
			et~al.}(2016)\citenamefont{Perez-Casta$\tilde{n}$eda, Jimenez-Riob\'oo, and
			Ramos}}]{perez2016}
	\bibinfo{author}{\bibfnamefont{T.}~\bibnamefont{Perez-Casta$\tilde{n}$eda}},
	\bibinfo{author}{\bibfnamefont{R.~J.} \bibnamefont{Jimenez-Riob\'oo}},
	\bibnamefont{and} \bibinfo{author}{\bibfnamefont{M.~A.} \bibnamefont{Ramos}},
	\bibinfo{journal}{Philos. Mag.} \textbf{\bibinfo{volume}{96}},
	\bibinfo{pages}{774} (\bibinfo{year}{2016}).
	
	\bibitem[{\citenamefont{Stone et~al.}(2001)\citenamefont{Stone, Hannon,
			Ishihara, Kitamura, Shirakawa, Sinclair, Umesaki, and
			Wright}}]{stone2001structure}
	\bibinfo{author}{\bibfnamefont{C.~E.} \bibnamefont{Stone}},
	\bibinfo{author}{\bibfnamefont{A.~C.} \bibnamefont{Hannon}},
	\bibinfo{author}{\bibfnamefont{T.}~\bibnamefont{Ishihara}},
	\bibinfo{author}{\bibfnamefont{N.}~\bibnamefont{Kitamura}},
	\bibinfo{author}{\bibfnamefont{Y.}~\bibnamefont{Shirakawa}},
	\bibinfo{author}{\bibfnamefont{R.~N.} \bibnamefont{Sinclair}},
	\bibinfo{author}{\bibfnamefont{N.}~\bibnamefont{Umesaki}}, \bibnamefont{and}
	\bibinfo{author}{\bibfnamefont{A.~C.} \bibnamefont{Wright}},
	\bibinfo{journal}{J. Non-Cryst. Solids} \textbf{\bibinfo{volume}{293}},
	\bibinfo{pages}{769} (\bibinfo{year}{2001}).
	
	\bibitem[{\citenamefont{R{\"u}ffer and Chumakov}(1996)}]{R}
	\bibinfo{author}{\bibfnamefont{R.}~\bibnamefont{R{\"u}ffer}} \bibnamefont{and}
	\bibinfo{author}{\bibfnamefont{A.~I.} \bibnamefont{Chumakov}},
	\bibinfo{journal}{Hyperfine Interact.} \textbf{\bibinfo{volume}{97}},
	\bibinfo{pages}{589} (\bibinfo{year}{1996}).
	
	\bibitem[{\citenamefont{Chumakov et~al.}(2009)\citenamefont{Chumakov, Bosak,
			and R\"uffer}}]{PhysRevB.80.094303}
	\bibinfo{author}{\bibfnamefont{A.~I.} \bibnamefont{Chumakov}},
	\bibinfo{author}{\bibfnamefont{A.}~\bibnamefont{Bosak}}, \bibnamefont{and}
	\bibinfo{author}{\bibfnamefont{R.}~\bibnamefont{R\"uffer}},
	\bibinfo{journal}{Phys. Rev. B} \textbf{\bibinfo{volume}{80}},
	\bibinfo{pages}{094303} (\bibinfo{year}{2009}).
	
	\bibitem[{\citenamefont{Carini~Jr et~al.}(2016)\citenamefont{Carini~Jr, Carini,
			Cosio, D'Angelo, and Rossi}}]{carini}
	\bibinfo{author}{\bibfnamefont{G.}~\bibnamefont{Carini~Jr}},
	\bibinfo{author}{\bibfnamefont{G.}~\bibnamefont{Carini}},
	\bibinfo{author}{\bibfnamefont{D.}~\bibnamefont{Cosio}},
	\bibinfo{author}{\bibfnamefont{G.}~\bibnamefont{D'Angelo}}, \bibnamefont{and}
	\bibinfo{author}{\bibfnamefont{F.}~\bibnamefont{Rossi}},
	\bibinfo{journal}{Philos. Mag.} \textbf{\bibinfo{volume}{96}},
	\bibinfo{pages}{761} (\bibinfo{year}{2016}).
	
	\bibitem[{\citenamefont{Monaco et~al.}(2006)\citenamefont{Monaco, Chumakov,
			Yue, Monaco, Comez, Fioretto, Crichton, and R\"uffer}}]{DOShyper}
	\bibinfo{author}{\bibfnamefont{A.}~\bibnamefont{Monaco}},
	\bibinfo{author}{\bibfnamefont{A.~I.} \bibnamefont{Chumakov}},
	\bibinfo{author}{\bibfnamefont{Y.~Z.} \bibnamefont{Yue}},
	\bibinfo{author}{\bibfnamefont{G.}~\bibnamefont{Monaco}},
	\bibinfo{author}{\bibfnamefont{L.}~\bibnamefont{Comez}},
	\bibinfo{author}{\bibfnamefont{D.}~\bibnamefont{Fioretto}},
	\bibinfo{author}{\bibfnamefont{W.~A.} \bibnamefont{Crichton}},
	\bibnamefont{and} \bibinfo{author}{\bibfnamefont{R.}~\bibnamefont{R\"uffer}},
	\bibinfo{journal}{Phys. Rev. Lett.} \textbf{\bibinfo{volume}{96}},
	\bibinfo{pages}{205502} (\bibinfo{year}{2006}).
	
	\bibitem[{\citenamefont{Singh et~al.}(2013)\citenamefont{Singh, Ediger, and
			de~Pablo}}]{singh_ultrastable_2013}
	\bibinfo{author}{\bibfnamefont{S.}~\bibnamefont{Singh}},
	\bibinfo{author}{\bibfnamefont{M.~D.} \bibnamefont{Ediger}},
	\bibnamefont{and} \bibinfo{author}{\bibfnamefont{J.~J.}
		\bibnamefont{de~Pablo}}, \bibinfo{journal}{Nat. Mater.}
	\textbf{\bibinfo{volume}{12}}, \bibinfo{pages}{139} (\bibinfo{year}{2013}).
	
	\bibitem[{\citenamefont{Orsingher et~al.}(2010)\citenamefont{Orsingher,
			Fontana, Gilioli, Carini~Jr, Carini, Tripodo, Unruh, and
			Buchenau}}]{BPGeO2dense}
	\bibinfo{author}{\bibfnamefont{L.}~\bibnamefont{Orsingher}},
	\bibinfo{author}{\bibfnamefont{A.}~\bibnamefont{Fontana}},
	\bibinfo{author}{\bibfnamefont{E.}~\bibnamefont{Gilioli}},
	\bibinfo{author}{\bibfnamefont{G.}~\bibnamefont{Carini~Jr}},
	\bibinfo{author}{\bibfnamefont{G.}~\bibnamefont{Carini}},
	\bibinfo{author}{\bibfnamefont{G.}~\bibnamefont{Tripodo}},
	\bibinfo{author}{\bibfnamefont{T.}~\bibnamefont{Unruh}}, \bibnamefont{and}
	\bibinfo{author}{\bibfnamefont{U.}~\bibnamefont{Buchenau}},
	\bibinfo{journal}{J. Chem. Phys.} \textbf{\bibinfo{volume}{132}},
	\bibinfo{pages}{124508} (\bibinfo{year}{2010}).
	
	\bibitem[{\citenamefont{Liu et~al.}(1995)\citenamefont{Liu, v.~L{\"o}hneysen,
			Weiss, and Arndt}}]{Liu1995}
	\bibinfo{author}{\bibfnamefont{X.}~\bibnamefont{Liu}},
	\bibinfo{author}{\bibfnamefont{H.}~\bibnamefont{v.~L{\"o}hneysen}},
	\bibinfo{author}{\bibfnamefont{G.}~\bibnamefont{Weiss}}, \bibnamefont{and}
	\bibinfo{author}{\bibfnamefont{J.}~\bibnamefont{Arndt}}, \bibinfo{journal}{Z.
		Phys. B} \textbf{\bibinfo{volume}{99}}, \bibinfo{pages}{49}
	(\bibinfo{year}{1995}).
	
	\bibitem[{\citenamefont{Ruffl\'e et~al.}(2006)\citenamefont{Ruffl\'e,
			Guimbreti$\grave{e}$re, Courtens, Vacher, and
			Monaco}}]{ruffle_glass-specific_2006}
	\bibinfo{author}{\bibfnamefont{B.}~\bibnamefont{Ruffl\'e}},
	\bibinfo{author}{\bibfnamefont{G.}~\bibnamefont{Guimbreti$\grave{e}$re}},
	\bibinfo{author}{\bibfnamefont{E.}~\bibnamefont{Courtens}},
	\bibinfo{author}{\bibfnamefont{R.}~\bibnamefont{Vacher}}, \bibnamefont{and}
	\bibinfo{author}{\bibfnamefont{G.}~\bibnamefont{Monaco}},
	\bibinfo{journal}{Phys. Rev. Lett.} \textbf{\bibinfo{volume}{96}},
	\bibinfo{pages}{045502} (\bibinfo{year}{2006}).
	
	\bibitem[{\citenamefont{Karpov et~al.}(1983)\citenamefont{Karpov, Klinger, and
			Ignat'ev}}]{Karpov1983}
	\bibinfo{author}{\bibfnamefont{V.~G.} \bibnamefont{Karpov}},
	\bibinfo{author}{\bibfnamefont{I.}~\bibnamefont{Klinger}}, \bibnamefont{and}
	\bibinfo{author}{\bibfnamefont{F.~N.} \bibnamefont{Ignat'ev}},
	\bibinfo{journal}{Zh. Eksp. Teor. Fiz} \textbf{\bibinfo{volume}{84}},
	\bibinfo{pages}{760} (\bibinfo{year}{1983}).
	
	\bibitem[{\citenamefont{Buchenau et~al.}(1992)\citenamefont{Buchenau, Galperin,
			Gurevich, Parshin, Ramos, and Schober}}]{SPM}
	\bibinfo{author}{\bibfnamefont{U.}~\bibnamefont{Buchenau}},
	\bibinfo{author}{\bibfnamefont{Y.~M.} \bibnamefont{Galperin}},
	\bibinfo{author}{\bibfnamefont{V.~L.} \bibnamefont{Gurevich}},
	\bibinfo{author}{\bibfnamefont{D.~A.} \bibnamefont{Parshin}},
	\bibinfo{author}{\bibfnamefont{M.~A.} \bibnamefont{Ramos}}, \bibnamefont{and}
	\bibinfo{author}{\bibfnamefont{H.~R.} \bibnamefont{Schober}},
	\bibinfo{journal}{Phys. Rev. B} \textbf{\bibinfo{volume}{46}},
	\bibinfo{pages}{2798} (\bibinfo{year}{1992}).
	
	\bibitem[{\citenamefont{Parshin et~al.}(2007)\citenamefont{Parshin, Schober,
			and Gurevich}}]{Parshin2007}
	\bibinfo{author}{\bibfnamefont{D.~A.} \bibnamefont{Parshin}},
	\bibinfo{author}{\bibfnamefont{H.~R.} \bibnamefont{Schober}},
	\bibnamefont{and} \bibinfo{author}{\bibfnamefont{V.~L.}
		\bibnamefont{Gurevich}}, \bibinfo{journal}{Phys. Rev. B}
	\textbf{\bibinfo{volume}{76}}, \bibinfo{pages}{064206}
	(\bibinfo{year}{2007}).
	
	\bibitem[{\citenamefont{Schober et~al.}(2014)\citenamefont{Schober, Buchenau,
			and Gurevich}}]{Schober2014}
	\bibinfo{author}{\bibfnamefont{H.~R.} \bibnamefont{Schober}},
	\bibinfo{author}{\bibfnamefont{U.}~\bibnamefont{Buchenau}}, \bibnamefont{and}
	\bibinfo{author}{\bibfnamefont{V.~L.} \bibnamefont{Gurevich}},
	\bibinfo{journal}{Phys. Rev. B} \textbf{\bibinfo{volume}{89}},
	\bibinfo{pages}{014204} (\bibinfo{year}{2014}).
	
	\bibitem[{\citenamefont{Schirmacher et~al.}(2008)\citenamefont{Schirmacher,
			Schmid, Tomaras, Viliani, Baldi, Ruocco, and
			Scopigno}}]{schirmacher2008vibrational}
	\bibinfo{author}{\bibfnamefont{W.}~\bibnamefont{Schirmacher}},
	\bibinfo{author}{\bibfnamefont{B.}~\bibnamefont{Schmid}},
	\bibinfo{author}{\bibfnamefont{C.}~\bibnamefont{Tomaras}},
	\bibinfo{author}{\bibfnamefont{G.}~\bibnamefont{Viliani}},
	\bibinfo{author}{\bibfnamefont{G.}~\bibnamefont{Baldi}},
	\bibinfo{author}{\bibfnamefont{G.}~\bibnamefont{Ruocco}}, \bibnamefont{and}
	\bibinfo{author}{\bibfnamefont{T.}~\bibnamefont{Scopigno}},
	\bibinfo{journal}{Phys. Status Solidi (C)} \textbf{\bibinfo{volume}{5}},
	\bibinfo{pages}{862} (\bibinfo{year}{2008}).
	
	\bibitem[{\citenamefont{Ruffl\'e et~al.}(2008)\citenamefont{Ruffl\'e, Parshin,
			Courtens, and Vacher}}]{PhysRevLett.100.015501}
	\bibinfo{author}{\bibfnamefont{B.}~\bibnamefont{Ruffl\'e}},
	\bibinfo{author}{\bibfnamefont{D.~A.} \bibnamefont{Parshin}},
	\bibinfo{author}{\bibfnamefont{E.}~\bibnamefont{Courtens}}, \bibnamefont{and}
	\bibinfo{author}{\bibfnamefont{R.}~\bibnamefont{Vacher}},
	\bibinfo{journal}{Phys. Rev. Lett.} \textbf{\bibinfo{volume}{100}},
	\bibinfo{pages}{015501} (\bibinfo{year}{2008}).
	
	\bibitem[{\citenamefont{Bove et~al.}(2008)\citenamefont{Bove, Petrillo,
			Fontana, and Sokolov}}]{doi:10.1063/1.2912060}
	\bibinfo{author}{\bibfnamefont{L.~E.} \bibnamefont{Bove}},
	\bibinfo{author}{\bibfnamefont{C.}~\bibnamefont{Petrillo}},
	\bibinfo{author}{\bibfnamefont{A.}~\bibnamefont{Fontana}}, \bibnamefont{and}
	\bibinfo{author}{\bibfnamefont{A.~P.} \bibnamefont{Sokolov}},
	\bibinfo{journal}{J. Chem. Phys.} \textbf{\bibinfo{volume}{128}},
	\bibinfo{pages}{184502} (\bibinfo{year}{2008}).
	
	\bibitem[{\citenamefont{Ruocco et~al.}(1999)\citenamefont{Ruocco, Sette,
			Di~Leonardo, Fioretto, Krisch, Lorenzen, Masciovecchio, Monaco, Pignon, and
			Scopigno}}]{PhysRevLett.83.5583}
	\bibinfo{author}{\bibfnamefont{G.}~\bibnamefont{Ruocco}},
	\bibinfo{author}{\bibfnamefont{F.}~\bibnamefont{Sette}},
	\bibinfo{author}{\bibfnamefont{R.}~\bibnamefont{Di~Leonardo}},
	\bibinfo{author}{\bibfnamefont{D.}~\bibnamefont{Fioretto}},
	\bibinfo{author}{\bibfnamefont{M.}~\bibnamefont{Krisch}},
	\bibinfo{author}{\bibfnamefont{M.}~\bibnamefont{Lorenzen}},
	\bibinfo{author}{\bibfnamefont{C.}~\bibnamefont{Masciovecchio}},
	\bibinfo{author}{\bibfnamefont{G.}~\bibnamefont{Monaco}},
	\bibinfo{author}{\bibfnamefont{F.}~\bibnamefont{Pignon}}, \bibnamefont{and}
	\bibinfo{author}{\bibfnamefont{T.}~\bibnamefont{Scopigno}},
	\bibinfo{journal}{Phys. Rev. Lett.} \textbf{\bibinfo{volume}{83}},
	\bibinfo{pages}{5583} (\bibinfo{year}{1999}).
	
	\bibitem[{\citenamefont{Ferrante et~al.}(2013)\citenamefont{Ferrante,
			Pontecorvo, Cerullo, Chiasera, Ruocco, Schirmacher, and Scopigno}}]{natcom}
	\bibinfo{author}{\bibfnamefont{C.}~\bibnamefont{Ferrante}},
	\bibinfo{author}{\bibfnamefont{E.}~\bibnamefont{Pontecorvo}},
	\bibinfo{author}{\bibfnamefont{G.}~\bibnamefont{Cerullo}},
	\bibinfo{author}{\bibfnamefont{A.}~\bibnamefont{Chiasera}},
	\bibinfo{author}{\bibfnamefont{G.}~\bibnamefont{Ruocco}},
	\bibinfo{author}{\bibfnamefont{W.}~\bibnamefont{Schirmacher}},
	\bibnamefont{and} \bibinfo{author}{\bibfnamefont{T.}~\bibnamefont{Scopigno}},
	\bibinfo{journal}{Nat. Commun.} \textbf{\bibinfo{volume}{4}},
	\bibinfo{pages}{1793} (\bibinfo{year}{2013}).
	
	\bibitem[{\citenamefont{Pogna et~al.}(2015)\citenamefont{Pogna,
			Rodr\'iguez-Tinoco, Cerullo, Ferrante, Rodr\'iguez-Viejo, and
			Scopigno}}]{Pogna2}
	\bibinfo{author}{\bibfnamefont{E.~A.~A.} \bibnamefont{Pogna}},
	\bibinfo{author}{\bibfnamefont{C.}~\bibnamefont{Rodr\'iguez-Tinoco}},
	\bibinfo{author}{\bibfnamefont{G.}~\bibnamefont{Cerullo}},
	\bibinfo{author}{\bibfnamefont{C.}~\bibnamefont{Ferrante}},
	\bibinfo{author}{\bibfnamefont{J.}~\bibnamefont{Rodr\'iguez-Viejo}},
	\bibnamefont{and} \bibinfo{author}{\bibfnamefont{T.}~\bibnamefont{Scopigno}},
	\bibinfo{journal}{Proc. Natl. Acad. Sci. USA} \textbf{\bibinfo{volume}{112}},
	\bibinfo{pages}{2331} (\bibinfo{year}{2015}).
	
	\bibitem[{\citenamefont{Kearns et~al.}(2010)\citenamefont{Kearns, Still, Fytas,
			and Ediger}}]{kearns2010}
	\bibinfo{author}{\bibfnamefont{K.~L.} \bibnamefont{Kearns}},
	\bibinfo{author}{\bibfnamefont{T.}~\bibnamefont{Still}},
	\bibinfo{author}{\bibfnamefont{G.}~\bibnamefont{Fytas}}, \bibnamefont{and}
	\bibinfo{author}{\bibfnamefont{M.~D.} \bibnamefont{Ediger}},
	\bibinfo{journal}{Adv. Mater.} \textbf{\bibinfo{volume}{22}},
	\bibinfo{pages}{39} (\bibinfo{year}{2010}).
	
	\bibitem[{\citenamefont{G\"otze and Mayr}(2000)}]{goetzePRE}
	\bibinfo{author}{\bibfnamefont{W.}~\bibnamefont{G\"otze}} \bibnamefont{and}
	\bibinfo{author}{\bibfnamefont{M.~R.} \bibnamefont{Mayr}},
	\bibinfo{journal}{Phys. Rev. E} \textbf{\bibinfo{volume}{61}},
	\bibinfo{pages}{587} (\bibinfo{year}{2000}).
	
	\bibitem[{\citenamefont{Scopigno et~al.}(2002)\citenamefont{Scopigno, Balucani,
			Ruocco, and Sette}}]{Scopigno2002341}
	\bibinfo{author}{\bibfnamefont{T.}~\bibnamefont{Scopigno}},
	\bibinfo{author}{\bibfnamefont{U.}~\bibnamefont{Balucani}},
	\bibinfo{author}{\bibfnamefont{G.}~\bibnamefont{Ruocco}}, \bibnamefont{and}
	\bibinfo{author}{\bibfnamefont{F.}~\bibnamefont{Sette}},
	\bibinfo{journal}{Physica B: Condens. Matter} \textbf{\bibinfo{volume}{318}},
	\bibinfo{pages}{341} (\bibinfo{year}{2002}).
	
\end{thebibliography}

\begin{thebibliography}{13}
	\expandafter\ifx\csname natexlab\endcsname\relax\def\natexlab#1{#1}\fi
	\expandafter\ifx\csname bibnamefont\endcsname\relax
	\def\bibnamefont#1{#1}\fi
	\expandafter\ifx\csname bibfnamefont\endcsname\relax
	\def\bibfnamefont#1{#1}\fi
	\expandafter\ifx\csname citenamefont\endcsname\relax
	\def\citenamefont#1{#1}\fi
	\expandafter\ifx\csname url\endcsname\relax
	\def\url#1{\texttt{#1}}\fi
	\expandafter\ifx\csname urlprefix\endcsname\relax\def\urlprefix{URL }\fi
	\providecommand{\bibinfo}[2]{#2}
	\providecommand{\eprint}[2][]{\url{#2}}
	
	\bibitem[{\citenamefont{Menor-Salv\'an
			et~al.}(2010)\citenamefont{Menor-Salv\'an, Najarro, Velasco, Rosales, Tornos,
			and Simoneit}}]{ambra_quiS}
	\bibinfo{author}{\bibfnamefont{C.}~\bibnamefont{Menor-Salv\'an}},
	\bibinfo{author}{\bibfnamefont{M.}~\bibnamefont{Najarro}},
	\bibinfo{author}{\bibfnamefont{F.}~\bibnamefont{Velasco}},
	\bibinfo{author}{\bibfnamefont{I.}~\bibnamefont{Rosales}},
	\bibinfo{author}{\bibfnamefont{F.}~\bibnamefont{Tornos}}, \bibnamefont{and}
	\bibinfo{author}{\bibfnamefont{B.~R.} \bibnamefont{Simoneit}},
	\bibinfo{journal}{Org. Geochem.} \textbf{\bibinfo{volume}{41}},
	\bibinfo{pages}{1089} (\bibinfo{year}{2010}).
	
	\bibitem[{\citenamefont{Navarro et~al.}(2010)\citenamefont{Navarro,
			Pe$\tilde{n}$alver, P\'erez-de La~fuente, Ortega-Blanco, Menor-Salv\'an,
			Barr\'on, Soriano, Rosales, L\'opez Del~Valle, Velasco et~al.}}]{ElSoplaoS}
	\bibinfo{author}{\bibfnamefont{M.}~\bibnamefont{Navarro}},
	\bibinfo{author}{\bibfnamefont{E.}~\bibnamefont{Pe$\tilde{n}$alver}},
	\bibinfo{author}{\bibfnamefont{R.}~\bibnamefont{P\'erez-de La~fuente}},
	\bibinfo{author}{\bibfnamefont{J.}~\bibnamefont{Ortega-Blanco}},
	\bibinfo{author}{\bibfnamefont{C.}~\bibnamefont{Menor-Salv\'an}},
	\bibinfo{author}{\bibfnamefont{E.}~\bibnamefont{Barr\'on}},
	\bibinfo{author}{\bibfnamefont{C.}~\bibnamefont{Soriano}},
	\bibinfo{author}{\bibfnamefont{I.}~\bibnamefont{Rosales}},
	\bibinfo{author}{\bibfnamefont{R.}~\bibnamefont{L\'opez Del~Valle}},
	\bibinfo{author}{\bibfnamefont{F.}~\bibnamefont{Velasco}},
	\bibnamefont{et~al.}, \bibinfo{journal}{Acta Geol. Sin. - Engl.}
	\textbf{\bibinfo{volume}{84}}, \bibinfo{pages}{959} (\bibinfo{year}{2010}).
	
	\bibitem[{\citenamefont{Tool}(1946)}]{tulS}
	\bibinfo{author}{\bibfnamefont{A.~Q.} \bibnamefont{Tool}}, \bibinfo{journal}{J.
		Am. Ceram. Soc.} \textbf{\bibinfo{volume}{29}}, \bibinfo{pages}{240}
	(\bibinfo{year}{1946}).
	
	\bibitem[{\citenamefont{Moynihan et~al.}(1976)\citenamefont{Moynihan, Easteal,
			and De~Bolt}}]{moyS}
	\bibinfo{author}{\bibfnamefont{C.~T.} \bibnamefont{Moynihan}},
	\bibinfo{author}{\bibfnamefont{A.~J.} \bibnamefont{Easteal}},
	\bibnamefont{and} \bibinfo{author}{\bibfnamefont{M.~A.}
		\bibnamefont{De~Bolt}}, \bibinfo{journal}{J. Am. Ceram. Soc.}
	\textbf{\bibinfo{volume}{59}}, \bibinfo{pages}{12} (\bibinfo{year}{1976}).
	
	\bibitem[{\citenamefont{Guo et~al.}(2011)\citenamefont{Guo, Potuzak, Mauro,
			Allan, Kiczenski, and Yue}}]{Guo20113230S}
	\bibinfo{author}{\bibfnamefont{X.}~\bibnamefont{Guo}},
	\bibinfo{author}{\bibfnamefont{M.}~\bibnamefont{Potuzak}},
	\bibinfo{author}{\bibfnamefont{J.~C.} \bibnamefont{Mauro}},
	\bibinfo{author}{\bibfnamefont{D.~C.} \bibnamefont{Allan}},
	\bibinfo{author}{\bibfnamefont{T.}~\bibnamefont{Kiczenski}},
	\bibnamefont{and} \bibinfo{author}{\bibfnamefont{Y.}~\bibnamefont{Yue}},
	\bibinfo{journal}{J. Non-Cryst. Solids} \textbf{\bibinfo{volume}{357}},
	\bibinfo{pages}{3230 } (\bibinfo{year}{2011}).
	
	\bibitem[{\citenamefont{Perez-Casta$\tilde{n}$eda
			et~al.}(2014)\citenamefont{Perez-Casta$\tilde{n}$eda, Jim\'enez-Riob\'oo, and
			Ramos}}]{perezS}
	\bibinfo{author}{\bibfnamefont{T.}~\bibnamefont{Perez-Casta$\tilde{n}$eda}},
	\bibinfo{author}{\bibfnamefont{R.~J.} \bibnamefont{Jim\'enez-Riob\'oo}},
	\bibnamefont{and} \bibinfo{author}{\bibfnamefont{M.~A.} \bibnamefont{Ramos}},
	\bibinfo{journal}{Phys. Rev. Lett.} \textbf{\bibinfo{volume}{112}},
	\bibinfo{pages}{165901} (\bibinfo{year}{2014}).
	
	\bibitem[{\citenamefont{Chumakov et~al.}(1996)\citenamefont{Chumakov, Baron,
			R\"uffer, Gr\"unsteudel, Gr\"unsteudel, and Meyer}}]{rufleINSS}
	\bibinfo{author}{\bibfnamefont{A.~I.} \bibnamefont{Chumakov}},
	\bibinfo{author}{\bibfnamefont{A.~Q.~R.} \bibnamefont{Baron}},
	\bibinfo{author}{\bibfnamefont{R.}~\bibnamefont{R\"uffer}},
	\bibinfo{author}{\bibfnamefont{H.}~\bibnamefont{Gr\"unsteudel}},
	\bibinfo{author}{\bibfnamefont{H.~F.} \bibnamefont{Gr\"unsteudel}},
	\bibnamefont{and} \bibinfo{author}{\bibfnamefont{A.}~\bibnamefont{Meyer}},
	\bibinfo{journal}{Phys. Rev. Lett.} \textbf{\bibinfo{volume}{76}},
	\bibinfo{pages}{4258} (\bibinfo{year}{1996}).
	
	\bibitem[{\citenamefont{Kohn and Chumakov}(2000)}]{Kohn2000S}
	\bibinfo{author}{\bibfnamefont{V.~G.} \bibnamefont{KohnS}} \bibnamefont{and}
	\bibinfo{author}{\bibfnamefont{A.~I.} \bibnamefont{Chumakov}},
	\bibinfo{journal}{Hyperfine Interact.} \textbf{\bibinfo{volume}{125}},
	\bibinfo{pages}{205} (\bibinfo{year}{2000}).
	
	\bibitem[{\citenamefont{Sette et~al.}(1998)\citenamefont{Sette, Krisch,
			Masciovecchio, Ruocco, and Monaco}}]{sette_dynamics_1998S}
	\bibinfo{author}{\bibfnamefont{F.}~\bibnamefont{Sette}},
	\bibinfo{author}{\bibfnamefont{M.~H.} \bibnamefont{Krisch}},
	\bibinfo{author}{\bibfnamefont{C.}~\bibnamefont{Masciovecchio}},
	\bibinfo{author}{\bibfnamefont{G.}~\bibnamefont{Ruocco}}, \bibnamefont{and}
	\bibinfo{author}{\bibfnamefont{G.}~\bibnamefont{Monaco}},
	\bibinfo{journal}{Science} \textbf{\bibinfo{volume}{280}},
	\bibinfo{pages}{1550} (\bibinfo{year}{1998}).
	
	\bibitem[{\citenamefont{Scopigno et~al.}(2003)\citenamefont{Scopigno, Ruocco,
			Sette, and Monaco}}]{scopigno_fragilityS}
	\bibinfo{author}{\bibfnamefont{T.}~\bibnamefont{Scopigno}},
	\bibinfo{author}{\bibfnamefont{G.}~\bibnamefont{Ruocco}},
	\bibinfo{author}{\bibfnamefont{F.}~\bibnamefont{Sette}}, \bibnamefont{and}
	\bibinfo{author}{\bibfnamefont{G.}~\bibnamefont{Monaco}},
	\bibinfo{journal}{Science} \textbf{\bibinfo{volume}{302}},
	\bibinfo{pages}{849} (\bibinfo{year}{2003}).
	
	\bibitem[{\citenamefont{Scopigno et~al.}(2010)\citenamefont{Scopigno,
			Cangialosi, and Ruocco}}]{scopigno_universal_2010S}
	\bibinfo{author}{\bibfnamefont{T.}~\bibnamefont{Scopigno}},
	\bibinfo{author}{\bibfnamefont{D.}~\bibnamefont{Cangialosi}},
	\bibnamefont{and} \bibinfo{author}{\bibfnamefont{G.}~\bibnamefont{Ruocco}},
	\bibinfo{journal}{Phys. Rev. B} \textbf{\bibinfo{volume}{81}},
	\bibinfo{pages}{100202} (\bibinfo{year}{2010}).
	
	\bibitem[{\citenamefont{Scopigno et~al.}(2005)\citenamefont{Scopigno, Ruocco,
			and Sette}}]{scopigno_microscopic_2005S}
	\bibinfo{author}{\bibfnamefont{T.}~\bibnamefont{Scopigno}},
	\bibinfo{author}{\bibfnamefont{G.}~\bibnamefont{Ruocco}}, \bibnamefont{and}
	\bibinfo{author}{\bibfnamefont{F.}~\bibnamefont{Sette}},
	\bibinfo{journal}{Rev. Mod. Phys.} \textbf{\bibinfo{volume}{77}},
	\bibinfo{pages}{881} (\bibinfo{year}{2005}).
	
	\bibitem[{\citenamefont{Perez-Casta$\tilde{n}$eda
			et~al.}(2016)\citenamefont{Perez-Casta$\tilde{n}$eda, Jim\'enez-Riob\'oo, and
			Ramos}}]{perez2016S}
	\bibinfo{author}{\bibfnamefont{T.}~\bibnamefont{Perez-Casta$\tilde{n}$eda}},
	\bibinfo{author}{\bibfnamefont{R.~J.} \bibnamefont{Jim\'enez-Riob\'oo}},
	\bibnamefont{and} \bibinfo{author}{\bibfnamefont{M.~A.} \bibnamefont{Ramos}},
	\bibinfo{journal}{Philos. Mag.} \textbf{\bibinfo{volume}{96}},
	\bibinfo{pages}{774} (\bibinfo{year}{2016}).
	
\end{thebibliography}

\end{document}